\documentclass[aps,prb,twocolumn,floatfix,superscriptaddress,longbibliography,showpacs,nofootinbib]{revtex4-1}

\usepackage[utf8]{inputenc}
\usepackage{graphicx}
\usepackage{dcolumn}
\usepackage{bm}
\usepackage{amsfonts}
\usepackage{amsmath}
\usepackage{amssymb}
\usepackage{braket}
\usepackage{color,soul}
\usepackage{wasysym}
\usepackage{mathrsfs}
\usepackage{times}
\usepackage[dvipsnames,svgnames,table]{xcolor}
\usepackage{soul}
\usepackage{hyperref}
\usepackage{fancyhdr}
\usepackage[english]{babel}
\usepackage{longtable}
\usepackage{multirow}
\usepackage{float}
\usepackage{comment}
\usepackage{units}
\usepackage{makecell}
\usepackage{enumitem}

\hypersetup{
    unicode=true,			
    pdftoolbar=true,			
    pdfmenubar=true,			
    pdffitwindow=false,			
    pdfstartview={FitH},		
    pdfauthor={FDR},			
    colorlinks=true,			
    linkcolor=NavyBlue,			
    citecolor=Maroon,			
    filecolor=NavyBlue,			
    urlcolor=NavyBlue       		
}

\makeatletter
\newsavebox{\@brx}
\newcommand{\llangle}[1][]{\savebox{\@brx}{\(\m@th{#1\langle}\)}\mathopen{\copy\@brx\kern-0.5\wd\@brx\usebox{\@brx}}}
\newcommand{\rrangle}[1][]{\savebox{\@brx}{\(\m@th{#1\rangle}\)}\mathclose{\copy\@brx\kern-0.5\wd\@brx\usebox{\@brx}}}
\makeatother

\newcommand{\new}[1]{\textcolor{black}{#1}}

\makeatother

\begin{document}
\title{Fluctuation-induced currents in suspended graphene nanoribbons: Adiabatic quantum pumping approach}

\author{Federico D. Ribetto}
\affiliation{Instituto de Física Enrique Gaviola (CONICET) and FaMAF, Universidad Nacional de Córdoba, Argentina}
\affiliation{Departamento de Física, Universidad Nacional de Río Cuarto, Río Cuarto, Argentina}

\author{Silvina A. Elaskar}
\affiliation{Facultad de Ciencias Químicas, Universidad Nacional de Córdoba, Argentina}

\author{Hernán L. Calvo}
\affiliation{Instituto de Física Enrique Gaviola (CONICET) and FaMAF, Universidad Nacional de Córdoba, Argentina}

\author{Raúl A. Bustos-Marún}
\email{Corresponding author: rbustos@famaf.unc.edu.ar}
\affiliation{Instituto de Física Enrique Gaviola (CONICET) and FaMAF, Universidad Nacional de Córdoba, Argentina}
\affiliation{Facultad de Ciencias Químicas, Universidad Nacional de Córdoba, Argentina}

\begin{abstract}
Graphene nanoribbons (GNRs) are thin strips of graphene with unique properties due to their structure and nanometric dimensions. They stand out as basic components for the construction of different types of nanoelectromechanical systems (NEMS), including some very promising sensors and pumps. However, various phenomena, such as unintended mechanical vibrations, can induce undesired electrical currents in these devices. Here, we take a quantum mechanical approach to analyze how currents induced by fluctuations (either thermal or of some other kind) in suspended GNRs contribute to the electric current. In particular, we study the pumping current induced by the adiabatic variation of the Hamiltonian
of the system when a transverse vibration (flexural mode) of a GNR suspended over a gate is excited. Our theoretical approach and results provide useful tools and rules of thumb to understand and control the charge current induced by fluctuations in GNR-based NEMS, which is important for their applications in nanoscale sensors, pumps, and energy harvesting devices.
\end{abstract}
\maketitle

\section{Introduction}

\begin{figure}[ht]
\centering \includegraphics[width=3.3in]{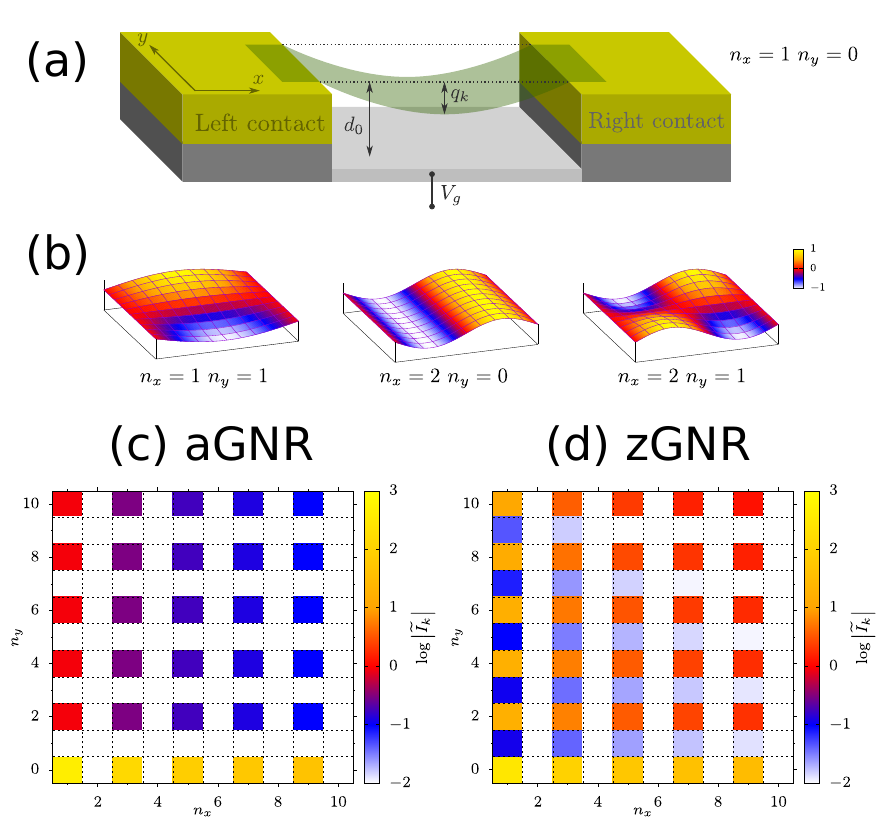} \caption{(a) Scheme of the studied system: A suspended GNR between two metallic contacts freely oscillating, at a distance $d_0$, over a gate with gate voltage $V_g$. We only consider the lowest frequency (acoustic) modes for the transversal vibration of the GNRs in a classical approximation and assume a rectangular membrane model to describe them. The shown mode corresponds to $(n_x,n_y)=(1,0)$. (b) Other examples of normal modes with (from left to right): $(n_x,n_y)=(1,1)$, $(2,0)$, and $(2,1)$. (c) and (d) Logarithm of the scaled emissivity $\tilde{I}_k$ in the low temperature limit as a function of $n_x$ and $n_y$ for square nanoribbons with armchair (aGNR) and zigzag (zGNR) edges. $\tilde{I}_k$ has units of 1/eV and is proportional to the maximum value of the pumping current during a period. The size of the
ribbons is $L=449\,a$ (with $a=0.246$~nm) and the Fermi energy is $\varepsilon_\text{F}=0.1$~eV.}
\label{fig:gnr-grillas}
\end{figure}

Being one of the strongest materials ever tested~\cite{lee2008} with large thermal and electrical conductivities,~\cite{chen2008,chen2011} graphene and, in particular, the more tunable graphene nanoribbons (GNRs) have become the conductive building blocks of innumerable nanoelectromechanical systems (NEMS). GNR-based NEMS have been studied as electromechanical resonators;~\cite{bunch2007,garcia2008,chen2009,croy2013} electron pumps;~\cite{zhu2009,prada2009,foa2011,sanjose2011,low2012,ingaramo2013,connolly2013} sensors of mass, pressure, strain, and temperature,~\cite{zang2015,davaji2017,khan2017} as well as detectors of vibrations~\cite{morenogarcia2022} and gases~\cite{gupta2021}; switches;~\cite{liao2011} ultrasmall accelerometers;~\cite{fan2018} and even viral detectors for COVID-19.~\cite{benshimon2022}

In all of the above examples, it is essential to study the various phenomena that can limit the use of the proposed devices, such as the noise in electrical currents. Besides the thermal (or Nyquist-Johnson) noise,~\cite{johnson1928,nyquist1928}
other phenomena can also interfere with the detected electrical currents. For example, suspended GNRs are in constant mechanical motion at room temperature, which can induce measurable oscillations in the electrical current, as recently shown.~\cite{thibado2020} Of course, oscillations are not only due to thermal excitations and other phenomena, such as the propagation of vibrational waves through the material, can also cause them.

Typically, the calculations of instantaneous electric currents induced by vibrations are based on classical models of time-dependent capacitances.~\cite{thibado2020} Here, instead, we adopt a quantum mechanical approach to this problem and analyze the contributions to the current due to the adiabatic quantum pumping~\cite{brouwer1998,bode2011,bustos2018} arising from the stochastic oscillation of GNRs.
Quantum pumping currents are a consequence of the delayed response of electronic wave functions to a time-dependent Hamiltonian which, in the present case, originates from the movement of the nuclei. In this regard, a comment is in order to avoid confusions. Since the oscillations of individual vibrational modes are independent, the generation of pumping currents produced by them is expected to average to zero over a period. This is because, to have a finite pumped charge per cycle, more than one time-dependent parameter needs to be moved with a fixed phase.~\cite{brouwer1998} However, even if the pumping currents have a null mean value, this does not prevent them from contributing instantaneously to the total current and, thus, to the current noise, when the Hamiltonian parameters move stochastically. This crucial point marks a clear difference with some previous works, e.g., Refs.~[\onlinecite{low2012,foa2011,ingaramo2013,connolly2013}]. While there the authors study the pumped charge in GNRs (meaning the average DC current per cycle), here we are interested in the maximum value of the instantaneous adiabatic pumping currents induced by stochastic fluctuations of GNRs.

In particular, we study a common configuration of GNR-based NEMS consisting of a GNR suspended over a controllable gate, see Fig.~\ref{fig:gnr-grillas}(a). There, we evaluate the contribution to the electric current due to quantum pumping induced by the movement of the lowest frequency (acoustic) modes for the transversal vibration of suspended GNRs, also known as flexural modes,~\cite{low2012} see Figs.~\ref{fig:gnr-grillas}(b) and (c). To this end, we adapted the theoretical description of adiabatic quantum pumping~\cite{brouwer1998} to the generic case of vibrational normal modes.~\cite{bustos2018} Our methodology and results are not only relevant for different types of sensors and
pumps based on GNRs, but they can also be extended to other applications such as energy harvesting at the nanoscale~\cite{kim2017,bustos2018,thibado2020} or to other systems such as carbon nanotubes.

The paper is organized as follows. In Sec.~\ref{sec:gnr-theory}, we present the theoretical framework, including the different models and approximations used in our calculations. In Sec.~\ref{sec:gnr-resultados}, we explore the role of different system's parameters on the fluctuation-induced currents, including the length, type of border, and Fermi energy of GNRs. At the end of this section we also present a realistic estimation of the expected value of currents. In Sec.~\ref{sec:gnr-noise} we develop a semiclassical theory to evaluate the zero-frequency noise of thermally induced pumping current and use it to compare with Nyquist-Johnson zero-frequency noise. Finally, in Sec.~\ref{sec:gnr-conclusions} we summarize and discuss the main results.

\section{Theoretical framework}
\label{sec:gnr-theory}

Due to the vast number of variables that can potentially influence the current induced by the oscillations of GNRs, throughout the manuscript we focus on certain limits, approximations, and simplified models that allow us to obtain simple expressions that nonetheless can be used to understand the general features of the studied phenomenon.

\subsection{GNR: electronic modelling}
\label{subsec:gnr-tb}

\textit{Hamiltonian.} \label{subsubsec:gnr-tb-grafeno} Considering energies close to the Fermi energy, the description of the electrons in graphene can be carried out by means of a tight-binding model.~\cite{foa2020} Given the energy difference between the $\sigma$ and $\pi$ molecular orbitals, and the fact that the band of $\pi$ orbitals is the one found around the Fermi energy, we will only consider $p_z$ atomic orbitals. Thus, we have the following Hamiltonian:
\new{
\begin{equation}
\hat{H} = \sum_i \varepsilon_i \hat{c}_{i}^\dag \hat{c}_i - \sum_{\braket{i,j}} t_{i,j} \hat{c}_{i}^\dag \hat{c}_j,
\label{hamiltonian-tb}
\end{equation}
where $\varepsilon_i$ is the energy of site $i$, $\hat{c}_i^\dag$ and $\hat{c}_i$ are the creation and annihilation operators in the $p_z$ orbital of site $i$, and $t_{i,j}$ represents the hopping amplitude between sites $i$ and $j$. Furthermore, the sum $\braket{i,j}$ is restricted to nearest neighbor sites. In the absence of defects and external disturbances, $\varepsilon_i$ is set to zero for simplicity, while the bare hopping amplitude takes the value $t_{i,j}\equiv t_0 = 2.66$~eV~\cite{foa2020}.}

\begin{figure}[t]
\includegraphics[width=2.2in]{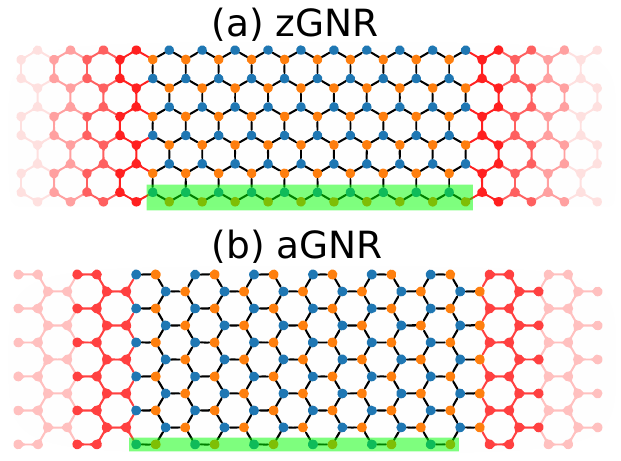} \caption{Examples of GNRs with zigzag (zGNR) and armchair (aGNR) edges (blue
and orange sites) connected to two semi-infinite contacts on the left and right (red sites). Note that the contacts are also made of GNRs of the same width and type. The green rectangles show how a row is defined for each type of GNR (to be used in Sec.~\ref{subsec:gnr-energia}).}
\label{fig:gnr-kwant-scheme}
\end{figure}

\textit{Electronic properties.} \label{subsubsec:gnr-tb-gnr} Graphene nanoribbons are usually cuts in a certain direction. Based on the direction of the cut, typically two edge types are described: \textit{zigzag} edge [Fig.~\ref{fig:gnr-kwant-scheme}(a)] and \textit{armchair} edge [Fig.~\ref{fig:gnr-kwant-scheme}(b)]. To classify the ribbons, the following convention will be used: GNRs with armchair (zigzag) edges are classified by the number of dimer lines (zigzag lines) across the width of the ribbon. In addition, the notation $N$-aGNR ($N$-zGNR) will be used for armchair (zigzag) GNRs, where $N$ is the number of dimer lines (zigzag lines).~\cite{foa2020}

$N$-aGNR and $N$-zGNR have very different electronic properties that arise from their contrasting boundary conditions. Some aGNR exhibit semiconductor behavior, while others are metallic. An analytical calculation of the eigenvalues of the tight-binding Hamiltonian allows us to show that the energy gap $\Delta_N$ of $N$-aGNR oscillates with the width of the ribbon,~\cite{cresti2008} besides the obvious decaying limit $\Delta_N \rightarrow 0$ for $N \rightarrow \infty$. In particular, we have that $\Delta_N=0$ for $N=3\ell+2$, where $\ell$ is an integer, making it metallic or semiconducting otherwise.~\cite{dutta2010,foa2020}

Unlike the previous case, the zGNRs retain the semimetallic character of graphene, regardless of their width. Another interesting feature is the formation of a pronounced peak in the density of states for $\varepsilon=0$, which results from the formation of partially flat and degenerate bands with zero energy.~\cite{foa2020} The presence of these highly confined electronic edge states has been confirmed by scanning tunneling microscopy and spectroscopy.~\cite{foa2020} They have a ``topological protection'' that makes them robust against different types of disturbances such as vacancies or Anderson-type noise~\cite{anderson1978,foa2020}.

\textit{System and contacts.} In our calculations we will assume the configurations shown in Fig.~\ref{fig:gnr-kwant-scheme} of the system depicted in Fig.~\ref{fig:gnr-grillas}(a). There, we define a finite central region (blue and orange dots in Fig.~\ref{fig:gnr-kwant-scheme}), that is able to oscillate and is coupled to two semi-infinite and identical contacts, left ($L$) and right ($R$) (red dots), that are also made of GNRs of the same type than that of the central region.

\subsection{GNR: Oscillating membrane model}
\label{subsec:gnr-vibraciones}

We are going to assume that the GNRs are large enough to be able to approximate its transversal normal modes of vibration by those of a rectangular membrane (limit of the continuum for low-frequency acoustic modes). We also assume that this membrane is placed in the $xy$ plane, and $z(x,y,t)$ is a function that describes its displacement with respect to said plane, see Fig.~\ref{fig:gnr-grillas}(a). Consequently, the equation that governs the transverse movement of the membrane is (see Appendix~\ref{app:modos} for more details) 
\begin{equation}
z(x,y,t) = \sum_{n_x,n_y=0}^\infty \sin\left(\frac{n_x\pi}{L_x}x\right)\cos\left(\frac{n_y\pi}{L_y}y\right) q_{n_x,n_y},\label{eq:gnr-modos}
\end{equation}
where $L_x$ and $L_y$ are the length and width of the central region, respectively, and the amplitude of the mode $q_{n_x,n_y}$
(defined by the pair of integers $n_x=1,2,...$ and $n_y=0,1,...$) is
\begin{equation}
q_{n_x,n_y} \equiv A_{n_x,n_y}\cos(\omega_{n_x,n_y}t+\phi_{n_x,n_y}).
\end{equation}
Fig.~\ref{fig:gnr-grillas}(b) shows some examples of modes for different pairs $(n_x,n_y)$. In the above equation, the phase
$\phi_{n_x,n_y}$ of the oscillation is arbitrary, the maximum amplitude of $q$ will be discussed afterwards, and the frequencies of the normal modes are given by
\begin{equation}
\omega_{n_x,n_y}=v\sqrt{\left(\frac{n_x\pi}{L_x}\right)^2+\left(\frac{n_y\pi}{L_y}\right)^2}, \label{eq:omega_nxny}
\end{equation}
where $v$ is the speed of sound in graphene (for low frequency transversal acoustic waves). For simplicity, from now on we will collapse the pair of indices $n_x$ and $n_y$ into a single one $k$. A central quantity that will be needed in our calculations is the maximum velocity of the normal modes, given by
\begin{equation}
\dot{q}_{k,\text{max}} = A_k\omega_k.
\end{equation}
The energy associated with the normal modes of graphene, see Appendix~\ref{app:mc}, can be written as 
\begin{equation}
E_k = m_\text{c} \left(\frac{1}{2}\dot{q}_k^2+\frac{1}{2}\omega_k^2 q_k^2\right) = \frac{1}{2} m_\text{c} A_k^2 \omega_k^2, \label{eq:E_k}
\end{equation}
where $m_\text{c}$ is the mass of a carbon atom. Now we can use equipartition theorem together with the above equation to derive an estimation for $A_k$, which yields
\begin{equation}
A_k=\frac{1}{\omega_k}\sqrt{\frac{2k_\text{B}T}{m_\text{c}}}.
\end{equation}
As we will see in Eq.~(\ref{eq:gnr-Irmax}), the above expression implies that the maximum value of the pumping current is independent of the frequency of the mode.

\subsection{Quantum pumping current}
\label{subsec:gnr-corriente}

The objective of this section is to derive an expression that allows us to calculate the maximum value of the pumping current on the reservoir $r$, $I_{r,k,\text{max}}^{(1)} \equiv \max[I_{r,k}^{(1)}(t)]$, over a period of oscillation of a given normal mode $k$ of a suspended GNR. This pumping current, typically denoted with a superscript (1), is associated with the first order term of an adiabatic expansion of the observable current, and arises from the delayed electronic response of the system to the time-variation of some classical parameters\new{, see Appendix~\ref{app:adiabatic}.}
It has been thoroughly studied by using different formalisms~\cite{brouwer1998,splettstoesser2005,arrachea2006} and in a variety contexts (even outside quantum transport~\cite{huanan2019}) but only as a way of generating DC currents controlled by the movement of at least two system's parameters.

Since our interest here lies in stochastic variations of the system's parameters (the vibrational normal modes $q_k$), it is necessary to highlight that the associated pumping currents average to zero in one period of $q_k(t)$. However, this does not mean that $I_{r,k}^{(1)}(t)$ is always zero. On the contrary, vibration-induced currents can instantaneously (at a given time) contribute to the total current which, as discussed in the introduction, can potentially interfere with current measurements in sensors based on GNRs, for example.

We will work under a perturbative limit of the modes, i.e., small amplitudes of the oscillation, and thus $\bm{q}(t) \approx \bm{q}_0$, where $\bm{q}$ is the vector containing the amplitudes $q_k$ of all normal modes, and $\bm{q}_0 \equiv \bm{0}$ is the equilibrium position of the system.
\footnote{\new{In our model, the equilibrium position (no excitation of vibrational modes for the oscillation of GNRs) corresponds to a flat still membrane, equivalent to setting $\bm{q}_0=\bm{0}$}}
Therefore, we can approximate $I_r^{(1)}(t)$ as
\begin{equation}
I_r^{(1)}(t) = e \sum_k \frac{\text{d}n_r(\bm{q})}{\text{d}q_k} \dot{q}_k(t) \approx e \sum_k\left(\frac{\text{d}n_r}{\text{d}q_k}\right)_{\bm{q}_0}\dot{q}_k(t),
\label{eq:gnr-Ir}
\end{equation}
where $(\text{d}n_r/\text{d}q_k)_{\bm{q}_0}$ is the emissivity due to the $k$-th normal mode evaluated at $\bm{q}_0$. This emissivity can be written in terms of the scattering matrix $S$ of the system as~\cite{brouwer1998}
\begin{equation}
\left(\frac{\text{d}n_r}{\text{d}q_k}\right)_{\bm{q}_0} = \int \frac{\text{d}\epsilon}{2\pi}
\left(-\frac{\partial f}{\partial\epsilon}\right)\sum_{\alpha \in r,\beta}\text{Im}\left[\frac{\partial S_{\alpha\beta}}{\partial q_k} S_{\alpha\beta}^*\right]_{\bm{q}_0},
\label{eq:gnr-emissiv}
\end{equation}
where $f$ is the equilibrium Fermi function (we are assuming a zero bias voltage configuration), $\alpha$ and $\beta$ are channel indices of the contacts, and we require $\alpha \in r$, since we want to calculate the emissivity at contact $r$ to determine the associated quantum pumping current $I_r^{(1)}$. If we additionally take the limit of low temperatures, Eq.~(\ref{eq:gnr-emissiv}) reduces to 
\begin{equation}
\left(\frac{\text{d}n_r}{\text{d}q_k}\right)_{\varepsilon_\text{F},\bm{q}_0} = \sum_{\alpha \in r,\beta}\frac{1}{2\pi}\text{Im}\left[\frac{\partial S_{\alpha\beta}}{\partial q_k}S_{\alpha\beta}^*\right]_{\varepsilon_\text{F},\bm{q}_0}.
\label{eq:gnr-emissiv-2}
\end{equation}
It is interesting to note that the above equation holds even at finite temperatures if the right-hand side of Eq.~(\ref{eq:gnr-emissiv}) depends linearly on the energy around $\varepsilon_\text{F}$. Due to the approximation used, emissivities do not depend on time. Then, the maximum value of the $k$-th contribution to the pumping current $I_{r,k}^{(1)}$ (defined through $I_r^{(1)}(t) = \sum_k I_{r,k}^{(1)}$) is simply
\begin{equation}
I_{r,k,\text{max}}^{(1)}=\left|e\left(\frac{\text{d}n_r}{\text{d}q_k}\right)_{\varepsilon_\text{F},\bm{q}_0}\dot{q}_{k,\text{max}}\right|,
\label{eq:gnr-Irmax}
\end{equation}
where $\dot{q}_{k,\text{max}}$ is the maximum value of $\dot{q}_k$.

\subsection{Emissivities}
\label{subsec:gnr-emissivities}

The explicit expression of the emissivity depends on the Hamiltonian model used. Here, we assume small displacements in the direction perpendicular to the GNR plane, $z$. Thus, we can take a linear regime of the diagonal elements of the Hamiltonian,
\begin{equation}
H_{ii}=E_0+\left(\frac{\partial E}{\partial z}\right)\delta z_i,
\label{eq:gnr-Hplacas}
\end{equation}
where $\delta z_i=[z_i(x_i,y_i,\bm{q})-z_i(x_i,y_i,\bm{q}_0)]$, $E_0=\varepsilon_i=0$, and the factor $(\partial E/\partial z)$ is the same for all sites. Taking the limit of large sizes for the GNR, within which our model should behave like a classical parallel plate capacitor, it can be shown (see Appendix~\ref{app:dedz}) that this factor should satisfy the relation
\begin{equation}
\left(\frac{\partial E}{\partial z}\right) = -\epsilon_0 \frac{A_\text{site} V_g^2}{d_0^2},
\end{equation}
where $A_\text{site}$ is the area associated with each individual site, $V_g$ is the potential difference between the membrane and the gate (gate voltage), $d_0$ is the distance between the GNR and the gate, and $\epsilon_0$ is the vacuum permittivity.

\new{In GNRs, the hopping amplitudes in the Hamiltonian usually take an
exponential dependence on the distance between first
neighbors~\cite{pereira2009}
\begin{equation}
t_{i,j} = t_0 e^{-b\left(\frac{|\bm{r}_i-\bm{r}_j|}{a_\text{cc}}-1\right)},
\end{equation}}
where $a_\text{cc}$ is the equilibrium distance between neighboring atoms and $b$ is a constant that sets how much the hoppings change with the distance $|\bm{r}_i-\bm{r}_j|$. Here, the distance $|\bm{r}_i-\bm{r}_j|$ is a quantity that depends on the modes that are being excited, and can be expressed as
\new{\begin{equation}
|\bm{r}_i-\bm{r}_j|=\sqrt{a_\text{cc}^2 + \left(\delta z_i-\delta z_j\right)^2}.
\end{equation}}
Given that we are considering that the amplitude $A_k$ of the modes is small, we can approximate 
\new{\begin{equation}
t_{i,j} \approx t_0 \left[ 1 - \frac{b}{2} \left(\frac{\delta z_i-\delta z_j}{a_\text{cc}}\right)^2 \right].
\label{eq:gnr-Hsepar}
\end{equation}}
The Hamiltonian of the system can then be divided into two parts, $\bm{H} = \bm{H}^{(E)}+\bm{H}^{(V)}$, where $\bm{H}^{(E)}$
contains $\bm{H}\left(\bm{q_{0}}\right)$ and the
site energies dependence on $\bm{q}$,
\begin{equation}
H_{ij}^{(E)} = \begin{cases}
E_0+\left(\dfrac{\partial E}{\partial z}\right)\delta z_i & i=j\\
-t_0 & i=j\pm1,
\end{cases}
\end{equation}
and $\bm{H}^{(V)}$ contains only the hoppings dependence on $\bm{q}$
\new{\begin{equation}
H_{ij}^{(V)} =
\begin{cases}
0 & i=j\\
 \dfrac{t_0 b}{2}\left(\dfrac{\delta z_i-\delta z_j}{a_\text{cc}}\right)^2 & i=j\pm1.
\end{cases}
\end{equation}}
Then, using the Fisher-Lee formula one can prove (see Appendix~\ref{app:gnr-bombeo-V}) that the derivatives with respect to $q_k$ of the scattering matrix can be decomposed into
\begin{equation}
\frac{\partial\bm{S}}{\partial q_k}=\frac{\partial\bm{S}^{(E)}}{\partial q_k}+\frac{\partial\bm{S}^{(V)}}{\partial q_k},\label{eq:s-derivada}
\end{equation}
where $\bm{S}^{(E)}$ and $\bm{S}^{(V)}$ are the scattering matrices obtained from Hamitonians $\bm{H}^{(E)}$ and $\bm{H}^{(V)}$, respectively. It can be readily shown that, for small oscillations, $\partial\bm{S}^{(V)} / \partial q_k$
can be neglected, see Appendix~\ref{app:gnr-bombeo-V}. Using this, and noticing that $\bm{S}(\bm{q}_0)=\bm{S}^{(E)}(\bm{q}_0)$, the emissivity takes the form
\begin{equation}
\left(\frac{\text{d}n_r}{\text{d}q_k}\right)_{\varepsilon_\text{F},\bm{q}_0} \approx \left(\frac{\text{d}n_r^{(E)}}{\text{d}q_k}\right)_{\varepsilon_\text{F},\bm{q}_0},
\end{equation}
where $\text{d}n_{r}^{(E)}/\text{d}q_k$ is the emissivity calculated with Hamiltonian $\bm{H}^{(E)}$.

To simplify the analysis of the next sections, we define the \textit{scaled} emissivity, which removes from $\text{d}n_{r}^{(E)}/\text{d}q_{k}$ the parameters related to particularities of the studied system, i.e.,
\begin{equation}
\left(\frac{\text{d}\tilde{n}_r^{(E)}}{\text{d}q_k}\right) \equiv \left(-\epsilon_0\frac{A_\text{site}V_g^2}{d_0^2}\right)^{-1}\left(\frac{\text{d}n_r^{(E)}}{\text{d}q_k}\right).
\end{equation}
Finally, the maximum value of the quantum pumping current, given by the oscillation of a mode $q_k$ with energy $k_\text{B}T$, yields
\begin{equation}
I_{r,k,\text{max}}^{(1)}=\left|\frac{e \epsilon_0 A_\text{site} V_g^2}{d_0^2}\right|\sqrt{\frac{2k_\text{B}T}{m_\text{c}}}\left|\frac{\text{d}\tilde{n}_r^{(E)}}{\text{d}q_k}\right|_{\varepsilon_\text{F},\bm{q}_0}.
\label{eq:gnr-Irmax-3}
\end{equation}
This expression can be divided into three contributions: the first one from left to right is the scaling factor and it accounts for the gate voltage and the distance between the GNR and the gate; the second one is the amplitude factor and it accounts for the amplitude of the $k$ mode oscillation (in this case given only by the temperature), and the last one is the scaled emissivity which is independent of the other quantities. To simplify the notation, from now on the scaled emissivity will be denoted as $\tilde{I}_k$, i.e.,
\begin{equation}
\tilde{I}_k \equiv \left(\frac{\text{d}\tilde{n}_r^{(E)}}{\text{d}q_k}\right)_{\varepsilon_\text{F},\bm{q}_0},
\end{equation}
where we omitted the $r$ subindex since it is irrelevant for the present case with equal contacts and zero bias voltage. This quantity, which has units of one over energy, can be transformed into a true current simply by multiplying it by the scaling and amplitude factors, see Eq.~(\ref{eq:gnr-Irmax-3}).

\section{Pumping currents induced by oscillations of the GNR}
\label{sec:gnr-resultados}

The main quantity to be discussed in Secs.~\ref{subsec:gnr-modos}, \ref{subsec:gnr-energia}, and \ref{subsec:gnr-size}, and plotted in Figs.~\ref{fig:gnr-grillas}, \ref{fig:gnr-eps}, \ref{fig:gnr-filas}, and \ref{fig:gnr-size}, is the scaled emissivity $\tilde{I}_k$ in the low temperature limit, where $\tilde{I}_k$ is proportional to the maximum value of the pumping current during a cycle of the $k$-mode. In sections~\ref{subsec:gnr-modos}, \ref{subsec:gnr-energia}, and \ref{subsec:gnr-size} we mainly study the dependence of $\tilde{I}_k$ on different characteristics of the GNRs, while in section \ref{subsec:gnr-estimacion} we discuss a realistic estimation of the maximum value of the vibration-induced pumping current in typical GNRs. To carry out these tasks, we use the theory and models previously presented, combining them with the numerical tools provided by the KWANT package~\cite{groth2014} for the calculation of the scattering matrices. In the plots all energies are in eV while distances are in units of \new{$a\equiv \sqrt{3} a_\text{cc} = 0.246$ nm}, which is the length of the primitive vectors of the associated Bravais lattice.~\cite{foa2020}

\subsection{Mode dependence}
\label{subsec:gnr-modos}

Here we study the effect of the mode oscillation on the quantum pumping current through the calculation of $\tilde{I}_k$ for different values of $n_x$ (in the range [1,10]) and $n_y$ (in the range [0,10]), taking into account both zigzag and armchair  edges. For the calculations we consider a square membrane with length $L_x$ and width $L_y$ equal to $449 \,a \sim 0.1$ $\mu$m. This value of $L_y$ ensures that the aGNR is metallic (see Sec.~\ref{subsubsec:gnr-tb-gnr}) and that the system is large enough to validate the rectangular membrane model (Sec.~\ref{subsec:gnr-vibraciones}). Regarding the Fermi energy, the value $\varepsilon_\text{F} = 0.1$~eV was taken, which ensures that we are working far from edge states for the zigzag case, which can cause the pumping current to diverge (see Sec.~\ref{subsec:gnr-energia}).

The dependence of $\tilde{I}_k$ on the excited mode is illustrated in Fig.~\ref{fig:gnr-grillas}(c) for armchair edges and in Fig.~\ref{fig:gnr-grillas}(d) for zigzag edges. It is clear, in both cases, a relationship between the symmetry of the normal modes and the pumping current: If $n_x$ is even or if $n_y$ is odd, the pumped current is negligible. This behavior can be explained by analyzing the contribution of each atom $l$ to the total emissivity of a mode $k$, that is
\begin{equation}
\left(\frac{\text{d}\tilde{n}_r}{\text{d}q_k}\right)=\sum_l \left(\frac{\text{d}\tilde{n}_r}{\text{d}z_l}\right)\frac{\partial z_l}{\partial q_k},
\label{eq:dndq_atomos}
\end{equation}
where $(\text{d}\tilde{n}_r/\text{d}z_l)$ is the emissivity given by moving only atom $l$. If we now assume that the system is
infinite and that all atoms are equivalent, in the sense that their variation equally affects the scattering matrix, we can write 
\begin{equation}
\left(\frac{\text{d}\tilde{n}_r}{\text{d}q_k}\right) \approx \left(\frac{\text{d}\tilde{n}_r}{\text{d}z_l}\right)\sum_l\frac{\partial z_l}{\partial q_k}.
\end{equation}
Then, based on this assumption, we see that it is only the position of each atom [$z_l \equiv z_l(x_l,y_l,\bm{q})$] and the normal mode involved $q_k$ which determines its contribution to the total current. Thus, for those vibrations where $n_x$ is even or $n_y$ is odd, any contribution to the pumping current is canceled by a contribution of the opposite sign. It is
important to note that, even if the previous approximation is not completely valid (for example, due to the finite size of the system), it can still be expected in certain cases that the maximum value of the pumping current is zero. This occurs when the system presents a reflection symmetry with respect to a line that cuts the nanoribbon in half. In that case, for any pair of sites $i$ and $j$ in mirror positions with respect to a line parallel (or perpendicular) to the direction
of transport, the emissivities per atom must be equivalent, i.e.,
\begin{equation}
\left(\frac{\text{d}\tilde{n}_r}{\text{d}z_i}\right)=\left(\frac{\text{d}\tilde{n}_r}{\text{d}z_j}\right).
\end{equation}
Given that for $n_y$ odd ($n_x$ even), the following relation holds
\begin{equation}
\frac{\partial z_i}{\partial q_k}=-\frac{\partial z_j}{\partial q_k},
\end{equation}
it is clear that $\tilde{I}_{k}=0$.

As can be seen in Fig.~\ref{fig:gnr-grillas}, the above arguments hold quite well for even $n_x$ on both types of ribbons. This
is reasonable considering that the system is truly infinite in the $x$ direction. Let us recall that we are using as contacts semi-infinite nanoribbons of the same width and type as that of the suspended system. In this way, the pumping current is strictly zero for $n_x$ even, when the positions of the equivalent atoms coincide exactly with a reflection with respect to a line perpendicular to the direction of transport that cuts the system in half. The deviations from this are very small for the values of $L$ used and therefore the pumping currents are negligible, as can be seen. With respect to the $y$-direction,
the system is finite and, furthermore, for GNRs with zigzag edges there is no strict reflection symmetry along this direction, see Fig.~\ref{fig:gnr-kwant-scheme}(a). For this reason, deviations from our previous reasoning are expected, and this explains that some current (although orders of magnitude smaller) can be appreciated for odd $n_y$. GNRs with armchair edges, on the other hand, do present a reflection symmetry along the $y$ axis, see Fig.~\ref{fig:gnr-kwant-scheme}(b), so it is expected that the pumping currents are exactly zero for odd $n_y$. 

We verified that the same type of behavior shown in Figs.~\ref{fig:gnr-grillas}(c) and (d) is also repeated for different Fermi energies and other sizes of the GNR, as long as the ribbon is not too small. Another general characteristic of the studied system is that the lower frequency modes yield, by far, the largest contribution to the pumping current. In particular, the mode that generates the highest value of $\tilde{I}_k$ is the one with $n_x=1$ and $n_y=0$. This is reasonable since,
for this mode, the movement of all the sites contributes with the same sign to the pumping current and, thus, there are no cancellations. In Fig.~\ref{fig:gnr-grillas}(a) we depict the shape of this mode, while in Fig.~\ref{fig:gnr-grillas}(b) we show some other modes with negligible contributions to $\tilde{I}_k$. Based on the previous results, from now on we will only work with the mode with $n_x=1$ and $n_y=0$, denoting the scaled emissivity as $\tilde{I}_{(1,0)}$.

\subsection{Fermi energy dependence}
\label{subsec:gnr-energia}

\begin{figure}[!ht]
\centering \includegraphics[width=2.8in]{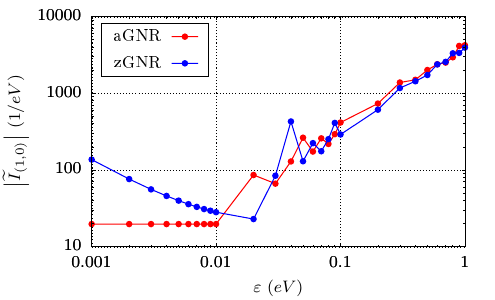} \caption{Scaled emissivity $\tilde{I}_{(1,0)}$ in units of 1/eV, as a function of the Fermi energy $\varepsilon_\text{F}$ for square GNRs with $L=449\,a$.}
\label{fig:gnr-eps}
\end{figure}

In this section we will study how the maximum pumping current behaves for different values of the Fermi energy, $\varepsilon_\text{F}$. Again, we will consider both edges for the graphene nanoribbon. In Fig.~\ref{fig:gnr-eps} we show $\tilde{I}_{(1,0)}$ as a function of the Fermi energy for square nanoribbons of sides $L_x=L_y=449\,a$ and for the mode with $n_x=1$ and $n_y=0$. In this figure it can be seen that for large Fermi energies the type of edge does not seem to have an effect on the maximum value of the pumping current. In both cases, the pumping current increases roughly linearly with
the number of conduction channels, which increases with the energy. For small Fermi energies, however, there is a clear difference in the behavior of the curves. In the case of aGNRs, $\tilde{I}_{(1,0)}$ converges to a fixed value as $\varepsilon_\text{F}$ goes to zero. Let us recall that because of the values of $L_y$ used, there is always one conduction channel at $\varepsilon_\text{F} \approx 0$ for this type of nanoribbon. On the other hand, for zGNRs, the maximum value of the current increases rapidly as $\varepsilon_\text{F}$ approaches zero. In fact, for $\varepsilon_\text{F}=0$, the pumping
current diverges, causing numerical problems in our calculations. This effect can be explained by noticing that zGNRs possess edge states at this energy with zero group velocity,~\cite{foa2020} causing the divergence of the associated density of states. The relation between the density of states and the pumped currents have been established in many different contexts.\cite{ingaramo2013,deghi2021}

\begin{figure}[!ht]
\centering \includegraphics[width=2.8in]{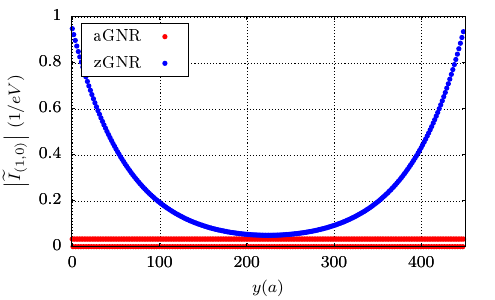} \caption{Contribution of each row of atoms [calculated using Eq.~(\ref{eq:dndq_atomos})] to the scaled emissivity $\tilde{I}_{(1,0)}$, for square GNRs of side $L=449\,a$, and $\varepsilon_\text{F}=0.001$~eV. The value of $y$ is the position of the row, in units of $a$. See Fig.~\ref{fig:gnr-kwant-scheme} for examples of rows for each type of GNR.}
\label{fig:gnr-filas}
\end{figure}

In order to confirm that edge states are causing the abrupt increase of the pumping currents, we evaluate the contribution of each row of atoms to the total current by means of Eq.~(\ref{eq:dndq_atomos}). In Fig.~\ref{fig:gnr-kwant-scheme} it can be seen how these rows are defined for each nanoribbon type (see the green area). Fig.~\ref{fig:gnr-filas} shows the contribution per row to $\tilde{I}_{(1,0)}$ for $\varepsilon_\text{F} = 0.001$~eV. As can be seen, for zGNR rows close to the edges (located at $y=0$ and $y=449$) are the ones that contribute the most to $\tilde{I}_{(1,0)}$. On the contrary, for aGNR rows near the edges do not seem to play any particular role in the currents. It should be mentioned that, in real systems, a true divergence of the pumping current is not expected since the coupling with the environment (causing for
example decoherence~\cite{pastawski2001}) should regularize it. However, at least a sharp peak of the pumping current should appear in potential experiments for $\varepsilon_\text{F}=0$. In our case, without a specific model for the interaction with the environment, we cannot estimate the value of the pumping current precisely at this value. Finally, it is interesting to discuss the effect of finite temperatures in Fig.~\ref{fig:gnr-eps}. Since the result of this is the convolution of the emissivity with the derivative of the Fermi-Dirac function [compare Eqs.~(\ref{eq:gnr-emissiv}) and (\ref{eq:gnr-emissiv-2})], its expected effect is simply to smooth (or average) the curves over a range of the order $k_\text{B}T$.
\footnote{\new{
Take into account that for Fermi energies $\varepsilon_\text{F}$ too close to the Dirac point 
($|\varepsilon_\text{F}| \lesssim k_\text{B}T$)
the calculation of pumping currents at finite temperatures would present numerical issues due to the divergence of emissivities there, see Eq.~(\ref{eq:gnr-emissiv})}}

According to the above discussion, there are two ways of increasing the pumped currents for zGNRs: either by adjusting the Fermi energy close to zero (and thus taking advantage of edge states and the divergence of the density of states that they cause), or by setting the Fermi energy to a large value (which naturally includes more conduction channels through which the current can be pumped). For aGNRs, the only available strategy is the second one.

\subsection{Size dependence}
\label{subsec:gnr-size}

In experiments with GNRs, these systems can sometimes have sizes on the order of a micrometer~\cite{bolotin2008,bolotin2008prl,singh2010,ma2013}. Since this involves a large number of atoms, performing numerical calculations for systems with those sizes involves an enormous computational cost. One way of estimating the value of the current $\tilde{I}_k$ for long nanoribbons is to study the behavior of this quantity as a function of size to make an extrapolation.

We calculate the maximum value of the pumping current for square nanoribbons of different sizes $L=L_x=L_y$ and for both types of edges. The particular widths used ($L_y$ values) were chosen such as to guaranty that the aGNRs are metallic. Based on the previous analysis, we took $n_x=1$ and $n_y=0$, and chose $\varepsilon_\text{F}=0.1$~eV for the calculations. This value of $\varepsilon_\text{F}$ is far enough from the Dirac point (thus avoiding numerical problems) and is, in principle, experimentally accessible.~\cite{zhang2020}

\begin{figure}[!ht]
\centering \includegraphics[width=2.8in]{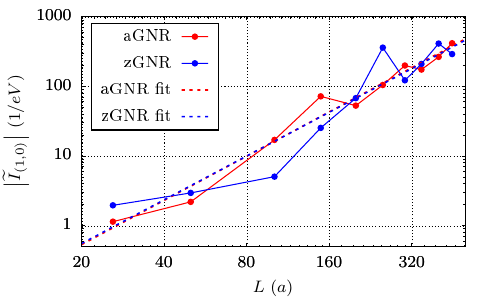} \caption{Scaled emissivity $\tilde{I}_{(1,0)}$ as a function of the system's size $L$ for square GNRs, with $\varepsilon_\text{F}=0.1$~eV. The linear fits in dotted lines, show the approximate quadratic dependence between $\tilde{I}_{(1,0)}$ and $L$. $\tilde{I}_{(1,0)} = 10^{-2.01} L^{2.11}$ for aGNR and $\tilde{I}_{(1,0)} = 10^{-2.97}L^{2.09}$ for zGNRs.}
\label{fig:gnr-size} 
\end{figure}

In Fig.~\ref{fig:gnr-size} we see, for both edges, that the log-log plot of $|\tilde{I}_k|$ with the size of the nanoribbons follows an approximate linear behavior for large values of $L$. There are also some dispersion around this linear behavior. However, this is not surprising since we are using the zero temperature limit of $\tilde{I}_k$, which is very sensitive to the density of states at the Fermi energy. Therefore, every time a new conduction channel appears at this energy (consequence of an increase in $L$), a jump in the density of states is expected (due to the emergence of van-Hove singularities) causing also a jump in $\tilde{I}_k$. The fittings shows that both edges follow an approximate quadratic dependence with $L$, for both edges. That is, the pumping currents scale with the surface of the system ($L^2$), and not with the width ($L$), as it is the case of the zero-order current $I^{(0)}$ (proportional to the number of conduction channels at the studied energy).

According to the above results, we can estimate the value of $\tilde{I}_k$ for a square nanoribbon with sides close to one micrometer, $L \approx 4065 \, a$. This gives an extrapolated scaled emissivity of approximately $\tilde{I}_k \sim 4 \times 10^{4} \, \text{eV}^{-1}$. This value will be useful in the following section.

\subsection{Realistic estimation of pumping currents}
\label{subsec:gnr-estimacion}

With the information collected so far, we are able to estimate the maximum value of the pumping current, $I_{L,(1,0),\text{max}}^{(1)}$, in a realistic context. Just for the sake of comparison, we will take the amplitude factor as given by temperature only, see Eq.~(\ref{eq:gnr-Irmax-3}).

For the reasons discussed in the previous sections we choose $n_x=1$, $n_y=0$, $\varepsilon_\text{F}=0.1$~eV and $L \approx 4065 \, a$. With these parameters, the extrapolated value of the scaled emissivity in the low temperature limit yields between $4.03 \times 10^5 \,\text{eV}^{-1}$ (extrapolating for aGNRs) and $3.74 \times 10^4 \, \text{eV}^{-1}$ (extrapolating for zGNRs). According to Eq.~(\ref{eq:gnr-Irmax-3}), we see that we still need to specify $d_0$, $T$, $V_g$ and $A_\text{site}$ to obtain a concrete value for the current. The factor $A_\text{site}$ is simply the area of the graphene unit cell divided by two, $A_\text{site}=0.026$ nm$^2$. For the rest of the parameters we took Ref.~[\onlinecite{bolotin2008}] as a guide ,
fixing the rest of the quantities at: $d_0=150$~nm, $T=5$~K, and $V_g=1$~V. 

By substituting all the above values in Eq.~(\ref{eq:gnr-Irmax-3}), we obtain a value of $I_{L,(1,0),\text{max}}^{(1)}$ between $1$ and $10$ pA. These contributions can be increased by taking advantage of the dependence of $I_{L,(1,0),\text{max}}^{(1)}$ on $d_0$. For example, reducing the separation to $d_0=15$ nm, it follows that $I_{L,(1,0),\text{max}}^{(1)}$ lies between $0.1$ and $1$ nA.

Just to put into context the above values of pumping currents we can compare them with some measured values of currents in nanodevices. In Ref.~[\onlinecite{thibado2020}], fluctuation-induced current of the order of 10 pA are observed working with micrometric sheets of suspended graphene at temperatures two orders of magnitude higher than that used in our estimate.
Another example is that of irradiated graphene experiments such as the one in Ref.~[\onlinecite{higuchi2017}], in which electric currents of the order of 10 pA at room temperature have also been measured. Working with mechanical resonators based on monolayer graphene, the authors of Ref.~[\onlinecite{chen2009}] measured currents of the order of pA at temperatures of 5 K.
In conclusion, the estimated values of pumping currents induced by thermal vibrations seem plausible of being measured and, in principle, their effect should not be neglected since they can potentially interfere in some experiments.

Before continuing, here a comment is in order. The formalism used to describe the pumping current assumes that the vibrational degree of freedom behaves classically. This is valid as long as the vibrational energy is far from the zero point energy, which for thermal excitation implies $\hbar\omega_k/2k_\text{B} \ll T$. Considering a GNR
with $L_x \sim 10^{-6}$ m, $n_x=1$, $n_y=0$, and taking $v \approx 12.9$ km/s,~\cite{cong2019} we estimate $\hbar\omega_{(1,0)}/2k_\text{B} \sim 0.05$ K. This very small temperature value also shows that the application of the equipartition theorem for the mechanical degrees of freedom used in Sec.~\ref{subsec:gnr-vibraciones} does not contradict the
low temperature limit for the energy dependence of the electronic scattering matrix used in Eq.~(\ref{eq:gnr-emissiv-2}).

Finally, it is important to highlight that the presented estimations correspond to pumping currents induced by thermal excitations of the vibrational modes. Other forms of ambient vibrations, such as sound waves traversing the sample, could give rise to considerably larger amplitudes of the normal modes [increasing the amplitude factor in Eq.~(\ref{eq:gnr-Irmax-3})] and thus much larger values of $I_{L,(1,0),\text{max}}^{(1)}$ are possible. Therefore, the above estimations are the minimal expected values of the pumping currents in a real case scenario.

\section{Current-noise induced by thermal oscillations of GNRs}
\label{sec:gnr-noise}

In previous sections, we show that fluctuation-induced pumping currents can be strong enough to be measurable, even for thermal excitations of the nanoribbons, affecting the observed currents of different nanodevices. Considering that stochastic fluctuations of the nanoribbons should induce what is essentially noise in the current determinations, it is fair to wonder how strong is this new form of noise, which we dubbed current-noise induced by thermal vibrations (CNITV), as compared with more standard forms of current noise. 

In this section, we develop a semiclassical approach to evaluate CNITV. This form of current-noise is associated with the first-order terms of the adiabatic expansion of the current $\hat{I}^{(1)}$ (or pumping current), and it should not be confused with Nyquist-Johnson or shot noise which are due to zero order terms of the current operator $\hat{I}^{(0)}$. In particular, we are wondering what is the correlation function of the pumping currents between contacts $\alpha$ and $\beta$, $S_{\alpha\beta}^{(1)}(\tau)$, induced by stochastic thermal variations of the parameters of a system's
Hamiltonian. Here, we evaluate this quantity by means of
\begin{equation}
S_{\alpha\beta}^{(1,\text{sc})}(\tau) = \Braket{ \left\{ \Delta I_{\alpha}^{(1)}\left(t\right)\Delta I_{\beta}^{(1)}\left(t'\right)\right\}_\text{s} }_\text{c},
\label{eq:S(t)-01}
\end{equation}
where $\tau=t-t'$, $\{A_\alpha A_\beta\}_\text{s} = (A_\alpha A_\beta + A_\beta A_\alpha)/2$,
$\braket{...}_\text{c}$ stands for the average over stochastic trajectories of the classical parameters of the Hamiltonian,
\footnote{\new{The electronic Hamiltonian $H$ varies when the GNR oscillates with a given normal mode (as explained in Sec.~\ref{subsec:gnr-emissivities}). Let us recall that the movement of the GNR is considered classically here, so $\bm{q}$ (the vector containing the amplitude of the normal modes) is a parameter of $H$. In our work, we are considering small amplitudes of the normal modes, so $\bm{q}(t)\approx \bm{q}_0$.
However, $\bm{\dot q}$ will still change stochastically due to thermal fluctuations. Therefore, when calculating the current noise one has to average over the stochastic variation of $\bm{\dot q}$, this is $\Braket{...}_\text{c}$, which is different from $\Braket{...}$, the expectation value of some quantity}}
and $\Delta I_\alpha^{(1)} = I_\alpha^{(1)}-\braket{I_\alpha^{(1)}}_\text{c}$. The pumping current $I_\alpha^{(1)}$ is the quantum expectation value $\braket{...}$ of the first order current operator $\hat{I}_\alpha^{(1)}$, i.e., $I_\alpha^{(1)}=\braket{\hat{I}_\alpha^{(1)}}$. This expectation value if zero once averaged over stochastic trajectories, i.e., $\braket{I_\alpha^{(1)}}_\text{c} = \llangle \hat{I}_\alpha^{(1)}\rrangle_\text{c}=0$. Therefore, $\Delta I_\alpha^{(1)}$ is simply $I_\alpha^{(1)}$.

\new{Note that Eq.~(\ref{eq:S(t)-01}) corresponds to a semiclassical approximation of the fully quantum current-noise~\cite{polianski2002,riwar2013}
\begin{equation}
S_{\alpha\beta}^{(1)}(\tau) = \llangle[\Big] \left\{ \Delta\hat{I}_\alpha^{(1)}(t)\Delta\hat{I}_\beta^{(1)}(t') \right\}_\text{s}\rrangle[\Big]_\text{c},\label{eq:S_quantum}
\end{equation}
where we replaced the current operators by their quantum expectation values.}
In this sense, at least for $\alpha=\beta$, Eq.~(\ref{eq:S(t)-01}) can be thought of as a lower limit to $S^{(1)}(\tau)$. First, note that their difference $\Delta S = S_{\alpha\alpha}^{(1)}(\tau)-S_{\alpha\alpha}^{(1,\text{sc})}(\tau)$
gives
\begin{equation}
\Delta S = \left\{ \llangle \hat{I}_\alpha^{(1)}(t)\hat{I}_\alpha^{(1)}(t') \rrangle_\text{c}  -\llangle \hat{I}_\alpha^{(1)}(t) \rangle \langle \hat{I}_\alpha^{(1)}(t')\rrangle_\text{c} \right\}_\text{s}.
\end{equation}
Now, for $\tau=0$, it holds 
\begin{equation}
\Braket{\left[ \hat{I}_\alpha^{(1)}(t)\right]^2} -\Braket{\hat{I}_\alpha^{(1)}(t)}^2 \geq 0,
\end{equation}
due to the Cauchy--Schwarz inequality, while for $\tau\rightarrow\infty$,
\begin{equation}
\llangle \hat{I}_\alpha^{(1)}(t)\hat{I}_\alpha^{(1)}(t')\rrangle_\text{c} = \llangle \hat{I}_\alpha^{(1)}(t) \rrangle_\text{c} = 0,
\end{equation}
since each trajectory of the classical parameters over which we are
averaging is stochastic. Therefore, $S_{\alpha\alpha}^{(1)} \ge S_{\alpha\alpha}^{(1,\text{sc})}$ for $\tau=0$ and $S_{\alpha\alpha}^{(1)} = S_{\alpha\alpha}^{(1,\text{sc})} = 0$ for $\tau \rightarrow \infty$.

In our case, the time-dependent variation of the acoustic transversal modes of graphene $q_i(t)$ is the origin of the current noise, and thus
\begin{equation}
\Delta I_\alpha^{(1)}(t) = e \sum_k \left( \frac{\text{d}n_\alpha}{\text{d}q_k}\right)\left(\dot{q}_k (t)-
\braket{\dot{q}_k(t)}_\text{c} \right).
\end{equation}
Replacing this in Eq.~(\ref{eq:S(t)-01}), taking into account that velocities of \new{different} modes are not correlated, i.e., $\braket{\dot{q}_k \dot{q}_{k'}}_\text{c}=\braket{\dot{q}_{k}}_\text{c}\braket{\dot{q}_{k'}}_\text{c}$, and that their stochastic average is zero, $\braket{\dot{q}_{k}(t)}_\text{c}=0$, we obtain
\begin{equation}
S_{\alpha\beta}^{(1,\text{sc})}(\tau) = e^2 \sum_k \left(\frac{\text{d}n_\alpha}{\text{d}q_k}\right)\!\left(\frac{\text{d}n_\beta}{\text{d}q_k}\right)\braket{\dot{q}_k(t)\dot{q}_k(t')}_\text{c}.
\label{eq:S(t)-02}
\end{equation}
\new{Now we will assume a generic model for the autocorrelation function of velocities
\begin{equation}
\braket{\dot{q}_k(t)\dot{q}_k(t')}_\text{c} \approx \frac{k_\text{B}T}{m_\text{c}} e^{-|t-t'|/\tau_k},\label{eq:<dotq_i_dotq_j>}
\end{equation}
which fulfills equipartition theorem for $t=t'$, while including the effects of a finite correlation time $\tau_k$ for each normal mode $k$. This kind of exponentially decaying correlation function can indeed be derived analytically for simple models, see for example Ref.~[\onlinecite{balakrishnan2021}].}
Inserting Eq.~(\ref{eq:<dotq_i_dotq_j>}) into Eq.~(\ref{eq:S(t)-02}) and taking the limit of small $\tau_k$ gives
\footnote{\new{This limit means that correlations decay instantaneously as compared with the measurement time. This would lead $S(\tau)$ to be proportional to the Dirac delta function (with a proportionality factor equal to the so called zero-frequency noise), which is an usual assumption in noise treatment.}}
\begin{equation}
S_{\alpha\beta}^{(1,\text{sc})}(\tau) \approx \frac{2e^2 k_\text{B}T}{m_\text{c}}
\left[\sum_k \left(\frac{\text{d}n_\alpha}{\text{d}q_k}\right)\!\left(\frac{\text{d}n_\beta}{\text{d}q_k}\right)\tau_k \right] \delta(\tau),
\end{equation}
where the term in brackets is the zero-frequency noise of the CNITV, $S_{\alpha\beta}^{(1,\text{sc})}(\omega=0)$. It is interesting to compare this term with the Nyquist-Johnson zero-frequency noise, $S_{\alpha\beta}^{(\text{NJ})}(\omega=0)$, typically used to assess current-noise. In the limit of small temperature, the latter is given as~\cite{blanter2000}
\begin{equation}
S_{\alpha\beta}^{(\text{NJ})}(\omega=0) = -4k_\text{B}T \frac{e^2}{h}\sum_{\eta\in\alpha,\eta'\in\beta} T_{\eta\eta'},
\end{equation}
where $T_{\eta\eta'}$ is the transmission coefficient between conduction channels $\eta$ and $\eta'$ belonging to contacts $\alpha$ and $\beta$, respectively, and $T_{\eta\eta} = -\sum_{\eta\neq\eta'}T_{\eta\eta'}$. If we wonder when $S_{\alpha\alpha}^{(1,\text{sc})}(\omega=0)$ will be the dominant source of current noise, i.e., $S_{\alpha\alpha}^{(1,\text{sc})}>S_{\alpha\alpha}^{(\text{NJ})}$, this implies the condition, in the low temperature limit,
\begin{equation}
\frac{h}{2m_\text{c}}\left(\frac{\partial E}{\partial z}\right)^2 \tilde{I}_{(1,0)}^2\tau_{(1,0)} \gtrsim -\sum_{\eta,\eta'\in\alpha} T_{\eta\eta'},
\end{equation}
where the right-hand side of the equation is the total transmittance and, in agreement with the previous sections, we assume that the mode $(n_x,n_y)=(1,0)$ gives the dominant contribution to the pumped current.

Typically, defects in quasi-one-dimensional systems like GNRs reduces dramatically the transmittance while pumping currents are not necessarily affected in the same way. Indeed, the peaks in the density of states caused by defects can favorably affect the pumping currents.~\cite{ingaramo2013} The scaling with the length of the sample also favors pumping currents. Note that the emissivity is approximately proportional to the length of the system [as Eq.~(\ref{eq:dndq_atomos}) suggests], while the total
transmittance in the coherent limit of the current is, at best, independent of the length of the sample. Let us recall that, in quasi-one-dimensional systems with defects, the transmittance is exponentially suppressed with the length.~\cite{cattena2010,fernandez2019}

The above reasoning sets some general conditions where one would expect the CNITV to be the dominant contribution to the current noise: elongated samples with defects. For example, we consider a GNR with $L_x = 800\,a$, $L_y=200\,a$ (which gives $\omega_{(1,0)} = 1.9 \times 10^{11}$ rad/s), 0.5\% of vacancies randomly distributed, $\varepsilon_\text{F}=0.1$~eV,
$d_0 = 15$~nm, and $V_g=1$~V. With these parameters, we obtained, after 70 independent runs giving different pairs of scaled emissivities and transmittances, that in 83\% of the samples $\tau_{(1,0)} = 1.05 \times 10^{-8}$ s is enough to make the CNITV the dominant contribution to current noise. This means that a quality factor (taking $Q=\tau_k \omega_k/2$) of only $10^3$ is required, which is a modest value compared with what is possible experimentally for graphene resonators at low temperatures.~\cite{takamura2016} Increasing vacancies up to 1\% makes that a quality factor of $10^3$ suffices for 92\% of the samples.

\section{Conclusions}
\label{sec:gnr-conclusions}

Motivated by the implementation of GNRs as pumps, sensors, and recent experiments on thermal fluctuation-induced currents, here
we developed a theoretical and numerical model to study the instantaneous value of quantum pumping currents in these systems. In particular, we focus on calculating the maximum value of the pumping current that arises from the excitation of the transverse acoustic modes of suspended GNRs. Although our calculations are strictly valid under certain conditions (namely classical limit of vibrations, small oscillations, low frequency transversal acoustic modes, large ribbon sizes, and the low temperature limit), our results can be helpful in estimating the magnitude of this type of currents. In this context, the developed theoretical tools and our results allow one to understand the role of the different parameters in the fluctuation-induced currents and assess under which conditions these currents can play an important role in actual experiments.

As particular results of the geometry studied, we found that: Pumping currents increase with the size of the system, being approximately proportional to the area of the ribbon and independent of the edge geometry for large sizes. The lowest frequency modes contribute the most to the pumping currents. The contribution of modes with $n_x$ even or $n_y$ odd is negligible, being strictly zero when the size of the system tends to infinity. While for large Fermi energies there are no significant differences between zigzag and armchair GNRs, for energies close to the Dirac point marked divergences appear. The behavior of aGNRs depends on their width, which makes them metallic or semiconductor. Metallic aGNRs present a relatively small and energy-independent pumping current, while semiconductor aGNRs have zero pumping current for Fermi energy close to the Dirac point. On the contrary, zGNRs present a strong increase of pumping currents close to the Dirac point due to the presence of their edge states.

Our estimations of GNRs' pumping currents in the studied geometries indicate that, in principle, even thermal fluctuations of the ribbons should produce measurable currents and that their effects should not be negligible for several experimental setups. It is worth noticing that other forms of excitation of the GNRs, such as propagating sound waves traversing the system, are potentially even more favorable for inducing pumping currents.

Finally, we developed a semiclassical theory to evaluate the current noise induced by thermal vibrations (CNITV). The comparison with Nyquist-Johnson noise allowed us to set some general conditions where one would expect the CNITV to be the dominant contribution to the current noise: elongated samples with defects.

\new{Our results, which include quantum effects, could provide a more accurate calculation of fluctuation-induced currents in experiments like the one in Ref.~[\onlinecite{thibado2020}]. Given that sensing is ultimately limited by the signal-to-noise ratio, our results can also help to improve the design of sensors based on suspended GNRs. This could be done, e.g., in optimal configurations that minimize current noise and/or maximize the signal of the fluctuation-induced current coming from a specific oscillation of the ribbon. If fluctuating currents arising from the oscillation of GNRs could be rectified (that would depend on the frequency and the magnitude of the current), then our results could be used to design energy harvesters that turn ambient mechanical excitations (above the thermal level) into electric power.}

\new{There are several further aspects that could be studied, including other vibrational modes, and different systems like carbon nanotubes or heterogeneous multilayer 2D materials. To assess the role of anharmonicities in the pumping currents, large oscillation amplitudes of the modes can be explored. Also, the effect of finite bias voltages on the induced current, or even the consequences of local voltage thermal fluctuations, give rise to interesting directions to follow.
In this regard, the present work paves the way to explore these and other exciting directions.}

\section{Acknowledgments}
This project was supported by Consejo Nacional de Investigaciones Cient\'ificas y T\'ecnicas (CONICET, PIP-2022-59241); Secretar\'ia de Ciencia y Tecnolog\'ia de la Universidad Nacional de C\'ordoba (SECyT-UNC, Proyecto Formar 2020); and Agencia Nacional de Promoción Científica y Tecnológica (ANPCyT, PICT-2018-03587).
We acknowledge the useful comments from Lucas Fernández-Alcázar and Sebastián Deghi. FDR, HLC, and RABM thank Luis E. F. Fo\`a Torres for fruitful discussions and hospitality during the 2023 ICTP conference at Universidad de Chile, Santiago de Chile.

\appendix

\section{Transverse normal modes of a rectangular elastic membrane}
\label{app:modos}

As mentioned in Sec.~\ref{subsec:gnr-vibraciones}, to study the vibrations in a GNR we will consider that the suspended system is large enough to be approximated by a rectangular elastic membrane. Assuming that the equilibrium position of the membrane defines the $xy$ plane, we will call $z(x,y,t)$ the displacement of the membrane with respect to said plane. This variable $z$ satisfies the wave equation
\begin{equation}
\nabla^2 z = \frac{1}{v^2}\frac{\partial^2 z}{\partial t^2}, 
\end{equation}
where $v$ is the speed of propagation of the wave. Making the replacement $z=F(x,y)q(t)$, we separate the above equation into a spatial equation (Helmholtz equation) and a temporal equation: 
\begin{equation}
(\nabla^2+k^2)F=0, \quad \ddot{q}+k^2 v^2 q=0. 
\end{equation}
The solutions to the temporal equation are of the form 
\begin{equation}
q(t)=A\cos(kvt)+B\sin(kvt), 
\end{equation}
while for the spatial part it is necessary to specify the boundary conditions. Once the corresponding eigenfunctions have been obtained, the general time-dependent solution, $z(x,y,t)$, can be constructed by superposition. Going back to the Helmholtz equation, it has the form
\begin{equation}
\frac{\partial^2 F}{\partial x^2}+\frac{\partial^2 F}{\partial y^2}+k^2 F = 0.
\label{helmholtz} 
\end{equation}
If the boundary conditions are separated in the $x$ and $y$ directions, we can apply separation of variables in the form $F(x,y)=X(x)Y(y)$. Substituting this into Eq.~(\ref{helmholtz}) and dividing by $F$ gives 
\begin{equation}
k^{2}+\frac{1}{X}\frac{\text{d}X^{2}}{\text{d}x^{2}}=-\frac{1}{Y}\frac{\text{d}Y^{2}}{\text{d}y^{2}}. 
\end{equation}
The left hand side is a function of $x$ only, while the right hand side is a function of $y$ only. Therefore, this relation is only valid if both sides are equal to a constant. With this in mind, we write
\begin{equation}
\frac{\text{d}X^2}{\text{d}x^2}=-k_x^2 X,
\label{eckx}
\end{equation}
and
\begin{equation}
\frac{\text{d}Y^{2}}{\text{d}y^2}=-k_y^2 Y,
\label{ecky}
\end{equation}
with $k_x$ and $k_y$ satisfying the relation $k_x^2+k_y^2=k^2$. Thus, the two-dimensional problem has been reduced to two one-dimensional problems, whose solutions are
\begin{align}
X(x) &= C_x \cos(k_x x) + D_x \sin(k_x x), \notag \\
Y(y) &= C_y \cos(k_y y) + D_y \sin(k_y y). 
\end{align}
We are now in position to specify the boundary conditions of the problem. We will think of a rectangular membrane with side $L_x$ on the $x$-axis and side $L_y$ on the $y$-axis, with the two edges in the $x$-direction fixed and the two edges in the $y$-direction free. This implies that the separate solutions of Eqs.~(\ref{eckx}) and (\ref{ecky}) must comply with
$X(0) = X(L_x) = 0$ and $Y'(0) = Y'(L_y)=0$, to ensure that the membrane does not move at the fixed edges and that
there is zero slope at the free edges. Applying these conditions, we arrive at
\begin{equation}
C_x = 0,\quad k_x =\frac{n_x\pi}{L_x},\quad n_x=0,1,2,\ldots,
\end{equation}
and
\begin{equation}
D_y = 0,\quad k_y =\frac{n_y\pi}{L_y},\quad n_y=0,1,2,\ldots,
\end{equation}
and in consequence:
\begin{equation}
X(x) = \sin(k_{x}x), \quad Y(y) = \cos(k_{y}y). 
\end{equation}
In conclusion, the solution of the spatial problem is of the form
\begin{equation}
F_{n_x,n_y}(x,y)=\sin\left(\frac{n_x\pi}{L_x}x\right)\cos\left(\frac{n_y\pi}{L_y}y\right), 
\end{equation}
with 
\begin{equation}
k_{n_x,n_y}^2=\frac{\omega_{n_x,n_y}^2}{v^2}=\left(\frac{n_x\pi}{L_x}\right)^2+\left(\frac{n_y\pi}{L_y}\right)^2. 
\end{equation}
The general solution to the two-dimensional wave equation can then be obtained by superposition of the normal modes, resulting in this case in Eq.~(\ref{eq:gnr-modos}) of the main text:
\begin{equation}
z(x,y,t)=\sum_{n_x=0}^\infty\sum_{n_y=0}^\infty \sin\left(\frac{n_x\pi}{L_x}x\right)\cos\left(\frac{n_y\pi}{L_y}y\right)q(t). 
\end{equation}

\section{Energy of normal modes}
\label{app:mc}

The total energy of a system composed of atoms labeled by $i$, and considering only the $z$ component of their respective
positions for simplicity, is
\begin{equation}
E_\text{tot} = \sum_i m_i \frac{\dot{z}_i^2}{2}+U_\text{tot}(\bm{z}), 
\end{equation}
where $m_i$ is the mass of the atom $i$ and $U_\text{tot}$ is the total potential energy of the system. As usual in any harmonic approximation we expand the potential up to second order in the displacement around the equilibrium position $\bm{z}_0$ (here sets equal to zero for simplicity). Using this, we obtain
\begin{equation}
E_\text{tot}-U_\text{tot}(\bm{z}_0) \approx \frac{m_\text{c}}{2} \left(\dot{\bm{z}}^\text{T}\dot{\bm{z}}+\frac{1}{m_\text{c}}\bm{z}^\text{T}\bm{\mathcal{H}}\bm{z}\right), 
\end{equation}
where $\bm{\mathcal{H}}$ is the Hessian matrix. Now, assuming all masses are equal ($m_i=m_\text{c}$) and inserting $\bm{U}^\text{T}\bm{U}=\bm{I}$, where $\bm{U}\bm{\mathcal{H}}\bm{U}^{T}/m_\text{c}=\bm{\Omega}_q$ with $\bm{\Omega}_q$ the diagonal matrix containing the square frequency of normal modes, the energy results in
\begin{equation}
E_\text{tot}-U_\text{tot}(\bm{z}_0) \approx \frac{m_\text{c}}{2} \left( \dot{\bm{z}}^\text{T}\bm{U}^\text{T}\bm{U}\dot{\bm{z}} 
+ \bm{z}^\text{T}\bm{U}^\text{T}\bm{\Omega}_q \bm{U} \bm{z} \right). 
\end{equation}
Finally, we recognize $\bm{q}=\bm{U}\bm{z}$ as the coordinate vector of the normal modes [which in our case we approximate as the $q_k$ of Eq.~(\ref{eq:gnr-modos})]. Then, the energy can be written as
\begin{equation}
E_\text{tot}-U_\text{tot}(\bm{z}_0) \approx \frac{m_\text{c}}{2} \left( \dot{\bm{q}}^\text{T}\dot{\bm{q}}+\bm{q}^\text{T}\bm{\Omega}_q \bm{q} \right) = \sum_k E_k, 
\end{equation}
with $E_k = m_\text{c} ( \dot{q}_k^2 + \omega_k^2 q_k^2)/2$ being the energy of the normal mode $k$.

\new{\section{Validity of the adiabatic approximation}
\label{app:adiabatic}}

\new{
Adiabatic quantum pumping arises from the first-order correction (in a frequency expansion) to the adiabatic approximation of an observable, the current in our case. Technical details of its derivation can be found among different contexts.~\cite{splettstoesser2006,arrachea2006,bode2011,deghi2021,ribetto2021}
In the present case, the adiabatic approximation and its first-order correction will be valid when the time it takes for the electrons to move along the lattice sites is much shorter than the time it takes for the  Hamiltonian to change. To assess this, we take a mode with some value $n_x$ of a GNR of length $L_x$ ($n_y=0$ for simplicity). The time $\tau_\text{e}$ electrons take to move (in a fluctuating GNR) between regions with the maximum and minimum values of the onsite energies, is $\tau_\text{e}= L_x/(2 n_{x}v_\text{F})$, where $v_\text{F}$ is the Fermi velocity. The frequency of the normal modes is given by Eq.~(\ref{eq:omega_nxny}), which sets the variation frequency of the electronic Hamiltonian. Taking $v_\text{F}=10^6$~m/s and $v_\text{T}=12.9\times10^3$~m/s (the velocity of sound for GNR transversal modes), yields a ratio between $\tau_\text{e}$ and the period of the vibration $\tau_\text{mech}=2\pi/\omega_{n_x,n_y}$ of $3.23\times10^{-3}$, which is a reasonably small value for the expansion. Remarkably, for $n_y=0$ this ratio is independent of the vibrational mode $n_x$, and given the expression for $\tau_\text{e}/\tau_\text{mech}$ we do not expect large variations from this ratio for small values of $n_y$.}

\section{Parallel plate model}
\label{app:dedz}

The goal of this appendix is to give an expression for the factor $\partial_z E$. The key idea is using the principle of correspondence between the quantum system described in Sec.~\ref{subsec:gnr-tb} and its classical analog taken as the capacitor formed by a continuous conductive membrane (the GNR) separated by a distance $d_0$ from the gate with a potential difference $V_g$. To simplify the derivation, we will assume that the GNR can only move as a whole in the $z$-direction, such that the classical analog is simply a parallel plate capacitor with separation $d_0$ between the plates. In this way, $z_i$
will be the same for all sites and the change in the onsite energies will also be the same, i.e. $\partial_z \bm{H}^{(E)} = (\partial_z E) \bm{I}$, with $\bm{I}$ the identity matrix. To simplify the notation, from now on we will omit the superscript ($E$) since we will be always working with the part of the Hamiltonian that represents the variations in the site energies of the system.
Now, the question that we want to answer is the following: What is the value of $\partial_z E$ in the tight-binding Hamiltonian
such that the system behaves as a parallel plate capacitor in the limit of large membranes?

Consider the energy of a capacitor with parallel plates of area $A$, a potential $V_g$ and separated by a distance $d$. The energy stored in this ``classical'' capacitor is 
\begin{equation}
E^{(\text{classic})}=\frac{1}{2}C V_g^2 = \frac{\epsilon_0 A}{2d} V_g^2.
\label{eq:Eclassic} 
\end{equation}
If we now think of small $z$-displacements from a given equilibrium position $d_0$, we can approximate the new energy as 
\begin{align}
E^{(\text{classic})} & \approx E_0^{(\text{classic})}+\left.\frac{\partial E^{(\text{classic})}}{\partial z}\right|_{d_0}z, \notag \\
 & = E_0^{(\text{classic})} + \left(-\frac{\epsilon_0 A}{2 d_0^2} V_g^2\right) z. 
\end{align}
Now considering the Hamiltonian model of the system, the total energy (near equilibrium) is
\begin{eqnarray}
\braket{\hat{H}} & = & \text{Tr}[\hat{H}\hat{\rho}^{(\text{eq})}]\nonumber \\
 & \approx & \braket{\hat{H}_0}+\text{Tr}[(\partial_{z}\hat{H})\hat{\rho}^{(\text{eq})}]z+\text{Tr}[\hat{H}\partial_{z}\hat{\rho}^{(\text{eq})}]z\nonumber \\
 & \approx & \braket{\hat{H}_{0}}+\left(\sum_{i}(\partial_{z}H_{ii})\rho_{ii}^{(\text{eq})}\right)z \nonumber \\
 &  & +\left(\sum_{ij}H_{ij}\partial_{z}\rho_{ji}^{(\text{eq})}\right)z,\label{eq:Equantum}
\end{eqnarray}
where we perform an expansion in $z$ around the equilibrium position
$z_{0}$. Here $\langle\hat{H}_{0}\rangle$ is the average energy
of the system at equilibrium ($z=0$). Note that $\hat{\rho}^{(\text{eq})}$
is diagonal in the energy basis, but not necessarily in the $i$ position
basis.

To find $\partial_z E$ we are going to demand that, in
the limit of large systems, the change in the total energy of the
system corresponds to that expected for the parallel plate capacitor
described above. This implies working Eq.~(\ref{eq:Equantum})
and then performing a comparison with Eq.~(\ref{eq:Eclassic}) in
the limit of macroscopic systems.

First, starting from Eq.~(\ref{eq:Equantum}), we will show that
the term with $\partial_{z}\rho_{ji}^{(\text{eq})}$ is zero. In order to easily calculate this amount, we are going to assume that
\begin{equation}
\hat{\rho}_{\text{tot}}^{(\text{eq})}=\hat{\rho}_{\text{leads}}^{(\text{eq})}\otimes\hat{\rho}_{\text{sys}}^{(\text{eq})},\quad\hat{\rho}_{\text{sys}}^{(\text{eq})}=\frac{\exp[-\beta(\hat{H}_{\text{sys}}-\mu\hat{N})]}{\Phi_{\text{sys}}}, 
\end{equation}
where $\Phi_{\text{sys}}=\text{Tr}[\exp(-\beta(\hat{H}-\mu\hat{N}))]$
and $\beta=1/k_{\text{B}}T$. Then, taking the Hamiltonian of the
system [see Eq. ~(\ref{eq:gnr-Hplacas})] and setting $H_{\text{sys}}\equiv H$,
we see that its derivatives are of the form
\begin{equation}
\partial_{z}\hat{H}=\left(\frac{\partial E}{\partial z}\right)\hat{1}\quad\Rightarrow\quad\partial_{z}\hat{H}^{n}=n\left(\frac{\partial E}{\partial z}\right)\hat{H}^{n-1}. 
\end{equation}
Similarly, it holds
\begin{eqnarray}
\partial_{z}(\hat{H}-\mu\hat{N}) & = & \left(\frac{\partial E}{\partial z}\right)\hat{1}, 
\end{eqnarray}
and then,
\begin{eqnarray}
\partial_{z}(\hat{H}-\hat{N})^{n} & = & n\left(\frac{\partial E}{\partial z}\right)(\hat{H}-\mu\hat{N})^{n-1}. 
\end{eqnarray}

Using these results, it is possible to calculate the derivative of the equilibrium density matrix 
\new{
\begin{eqnarray}
\partial_{z}\rho_{ji}^{(\text{eq})} & = & \langle j|\frac{\partial_{z}e^{-\beta(\hat{H}-\mu\hat{N})}}{\Phi}|i\rangle
 \notag \\ &&
  +\langle j|e^{-\beta(\hat{H}-\mu\hat{N})}\partial_{z}\left(\frac{1}{\Phi}\right)|i\rangle. 
\end{eqnarray}
}
Working with the first term, we get
\begin{eqnarray}
\langle j|\partial_{z}\left(\frac{e^{-\beta(\hat{H}-\mu\hat{N})}}{\Phi}\right)|i\rangle & = & (-\beta)\left(\frac{\partial E}{\partial z}\right)\rho_{ji}^{(\text{eq})}. 
\end{eqnarray}
On the other hand, working the second term we arrive at
\begin{eqnarray}
\langle j|e^{-\beta(\hat{H}-\mu\hat{N})}\partial_{z}\left(\frac{1}{\Phi}\right)|i\rangle & = & \beta\left(\frac{\partial E}{\partial z}\right)\rho_{ji}^{(\text{eq})}. 
\end{eqnarray}
Thus, joining the two previous results yields $\partial_{z}\rho_{ji}^{(\text{eq})}=0$.

The fact that $\partial_{z}\rho_{ji}^{(\text{eq})}=0$ and the assumption
$\partial_{z}H=(\partial E/\partial z)\bm{I}$ allows us to
write $\langle H\rangle$ as
\begin{equation}
\langle H\rangle\approx\langle H_{0}\rangle+\left(\frac{\partial E}{\partial z}\right)n_\text{sys}^{(\text{eq})}z, 
\end{equation}
where $n_\text{sys}^{(\text{eq})}=\left(\sum_{i}\rho_{ii}^{(\text{eq})}\right)$
is the mean number of particles within the system at equilibrium.
Therefore, for our Hamiltonian model to be consistent with macroscopic
theory, we must do 
\begin{eqnarray}
\underset{A\rightarrow\infty}{\lim}\left(-\frac{1}{2}\epsilon_0\frac{A}{d_{0}^{2}}V_g^{2}\right)z & = & \underset{A\rightarrow\infty}{\lim}\left(\frac{\partial E}{\partial z}\right)n_\text{sys}^{(\text{eq})}z, 
\end{eqnarray}
which gives
\begin{eqnarray}
\left(\frac{\partial E}{\partial z}\right) & = & \dfrac{-\dfrac{1}{2}\dfrac{\epsilon_0 V_g^{2}}{d_{0}^{2}}}{\underset{A\rightarrow\infty}{\lim}\dfrac{n_{\text{sys}}^{(\text{eq})}}{A}}. 
\end{eqnarray}
Note that $n_\text{sys}^{(\text{eq})}$
is the number of sites multiplied by $1/2$ (since two electrons enter
per site according to the Pauli exclusion principle), and that the
total area $A$ can be thought of as the number of sites multiplied
by the area occupied by each site, $A_\text{site}$.
Then we can write
\begin{equation}
\underset{A\rightarrow\infty}{\lim}\frac{n_\text{sys}^{(\text{eq})}}{A}=\frac{1}{2A_{\text{site}}}, 
\end{equation}
and in consequence,
\begin{equation}
\left(\frac{\partial E}{\partial z}\right)=-\epsilon_0 \frac{A_{\text{site}}V_g^{2}}{d_{0}^{2}}. 
\end{equation}

\section{Pumping currents due to coupling variations}
\label{app:gnr-bombeo-V}

In this appendix we will show that, in the limit of small oscillations
of flexural modes, the contribution to the pumping current due to
hopping variations between neighboring atoms can be neglected. More
specifically, this involves deriving Eq.~(\ref{eq:s-derivada}) and
proving that $\partial\bm{S}^{(V)} / \partial q_{k}$ tends to
zero in the perturbative limit. To carry out this task, we will use
the matrix form of the Fisher-Lee formula ~\cite{fisher1981,bode2012,bustos2018}
\begin{equation}
\bm{S}=\bm{I}-2i\bm{W}^{\dagger}\bm{G}^{R}\bm{W}.\label{eq:fisher_lee}
\end{equation}
Here, $\bm{G}^{R}$ is the retarded Green function given by 
\begin{equation}
\bm{G}^{R}=\underset{\eta\rightarrow0^{+}}{\lim}\left[(\varepsilon+i\eta)\bm{I}-\bm{H}-\bm{\Sigma}^{R}\right]^{-1}, 
\end{equation}
where $\bm{H}$ is the Hamiltonian of the system without the contacts,
$\bm{\Sigma}^{R}$ is the retarded self-energy due to the contacts,
and $\varepsilon$ is the energy of the electrons. The matrix $\bm{W}$
comes from the relation
\begin{equation}
\bm{\Gamma}_{\alpha}=\bm{W}^{\dagger}\bm{\Pi}_{\alpha}\bm{W}, 
\end{equation}
where $\bm{\Pi}_{\alpha}$ is the projection operator towards the
channel $\alpha$ of some reservoir $r$, and $\bm{\Gamma}_{\alpha}$
is the contribution, due to channel $\alpha$, of the imaginary part
of the self-energy $\bm{\Sigma}^{R}$, i.e., $\bm{\Gamma}=-\text{Im}(\bm{\Sigma}^{R})$
and $\bm{\Gamma}=\sum_{\alpha}\bm{\Gamma}_{\alpha}$.

Using Eq.~(\ref{eq:fisher_lee}) and $\partial\bm{G}^{r}=-\bm{G}^{r}\partial\left[\bm{G}^{r}\right]^{-1}\bm{G}^{r}$
we find
\begin{eqnarray}
\frac{\partial\bm{S}}{\partial q_{k}} & = & (-2i)\bm{W}^{\dagger}\bm{G}^{r}\frac{\partial\bm{H}^{(E)}}{\partial q_{k}}\bm{G}^{r}\bm{W}\nonumber \\
 &  & +(-2i)\bm{W}^{\dagger}\bm{G}^{r}\frac{\partial\bm{H}^{(V)}}{\partial q_{k}}\bm{G}^{r}\bm{W}  \notag \\
 & = & \frac{\partial\bm{S}^{(E)}}{\partial q_{k}}+\frac{\partial\bm{S}^{(V)}}{\partial q_{k}},  
\end{eqnarray}
proving Eq.~(\ref{eq:s-derivada}).

Now, using the chain rule
\new{
\begin{eqnarray}
\frac{\partial\bm{H}^{(V)}}{\partial q_{k}} & = & \sum_{\ell}\frac{\partial\bm{H}^{(V)}}{\partial z_{\ell}}\frac{\partial z_{\ell}}{\partial q_{k}}, 
\end{eqnarray}
}
where $z_{\ell}$ is the displacement of atom $\ell$ in the $z$
direction, and studying the matrix elements of $\partial\bm{H}^{(V)}/\partial z_{\ell}$
when the system is in equilibrium we arrive at that 
\new{\begin{equation}
\left(\frac{\partial H_{ij}^{(V)}}{\partial z_\ell}\right)_{\bm{q}_0} = \left[(z_i-z_j)\frac{t_0 b}{a_\text{cc}}\frac{\partial}{\partial z_\ell}(z_i-z_j)\right]_{\bm{q}_0}. 
\end{equation}}
Since evaluating at $\bm{q}_{0}$ implies taking $z_{i}=0$, $\forall i$,
we have that 
\begin{equation}
\left(\frac{\partial H_{ij}^{(V)}}{\partial z_{\ell}}\right)_{\bm{q}_{0}}\sim\left(z_{i}-z_{j}\right)_{\bm{q}_{0}}=0. 
\end{equation}
In short, for small oscillations around the equilibrium position,
the hopping variation does not contribute to the pumping current induced
by flexural modes.

\bibliographystyle{apsrev4-1_title}
\bibliography{cite_v1p3}

\end{document}